\documentstyle[12pt]{article}
\date{ }

\title{The St\"uckelberg formalism is a field-enlarging
transformation}

\author{ J. S\l adkowski $^{\dag}$  \\
Fakult\" at f\" ur   Physik Universit\"at Bielefeld,\\
D-4800 Bielefeld 1, Universit\"atsstrasse 25, Germany}
\begin{document}
\baselineskip9mm
\maketitle
\begin{abstract}
\baselineskip9mm

It is shown that the St\"uckelberg formalism can be regarded as a
field-enlarging transformation that introduces an additional gauge
symmetry to the considered model. The appropriate BRST charge can be
defined. The physical state condition, demanding that that a physical
state is to be anihilated by the BRST charge, is shown to be
equivalent to the St\"uckelberg condition. Several applications
of the new approach to the formalism are presented. The comparison
with the BFV procedure is given.

\end{abstract}
\vspace{20mm}

$^{\dag}$ A. von Humboldt Fellow; permanent address: Dept.
of Field Theory and
Particle Physics, University of Silesia, Pl 40007 Katowice, Poland.

\newpage
\section{Introduction}

\ \ \ The choice of variables used to describe a quantum field
theory should not have any physical significance. Such a field
redefinition invariance is a quite nontrivial problem in quantum
field theory. Complication may arise already at the level of free
theories. A well known complication arise when one considers
renormalizability of a gauge theory: one is forced to introduce
extra degrees of freedom to show it (the unitary gauge is
nonrenormalizable). Recently, it has been proposed to apply the BRST
symmetry idea to the field redefinition problem [1, 2]. We would
like to show how these tools work in the St\"uckelberg formalism
case [3]. The application of the general formalism presented in
[1, 2] to a concrete and popular physical model allows for a deeper
insight  into the St\"uckelberg formalism. Moreover, it suggests
its generalization to the general case of a vector field having
couplings of the non-Yang-Mills type. We shall also discuss the
relation of the field-enlarging transformation to the Batalin-
Fradkin-Vilkoviski formalism [4, 5] on the example of the
anomalous $U(1)$ chiral gauge theory [6-10].\\

\section{Abelian case}

\ \ \ Let us consider an Abelian massive gauge field $A_{\mu}$
with the following Lagrange density:

$${\it L}=-\frac{1}{4} F_{\mu \nu}F^{\mu \nu} + \frac{m^2}{2}
A_{\mu}A^{\mu} \ ,\eqno(1)$$
where

$$F_{\mu \nu} =\partial _{\mu}A_{\nu} - \partial _
{\nu}A_{\mu}$$
and apply to it the following field-enlarging transformation
[1,2,8,9]:

$$A_{\mu} =A'_{\mu} + \frac{1}{m} \partial_{\mu}\phi \equiv g_{\mu}
\left( A',\phi\right) \ . \eqno(2)$$
The substitution of (2) into (1) gives (we will write $A_{\mu}$
instead of $A'_{\mu}$)
$${\it L}=-\frac{1}{4} F_{\mu \nu}F^{\mu \nu} + \frac{m^2}{2}
A_{\mu}A^{\mu} + \frac{1}{2} \partial_{\mu}\phi\partial^
{\mu}\phi + mA_{\mu}\partial^{\mu}\phi \ .\eqno(3)$$
The above Lagrangian density is invariant with respect to the
following gauge transformations [1,2]:

$$\delta \phi \left( x \right) = \alpha \left( x \right) \eqno(4a)$$

$$\delta A_{\mu}\left( x \right) = -\int d^{4}zd^{4}y \left[
{{\delta g_{\mu}
\left( A,\phi \right)}\over{\delta A_{\nu}}}\right] ^{-1}
\left( x,y \right) {{\delta g_{\nu} \left( A,\phi \right)}\over
{\delta \phi}}
\left( y,z \right) \alpha \left( z \right)= - \frac{1}{m}\partial _
{\mu}\alpha \left( x \right) \ , \eqno(4b)$$
where $\alpha$ is an arbitrary function. In order to quantize this
we have to remove the gauge freedom. Let ua consider the following
gauge fixing term

$${\it L}_{gf} = - \lambda \left( \partial _{\mu}A^{\mu} - \frac
{m}{2\lambda}\phi \right)^{2} \ . \eqno(5)$$
The gauge-fixed Lagrangian density takes the form
$${\it L}=-\frac{1}{4} F_{\mu \nu}F^{\mu \nu} + \frac{m^2}{2}
A_{\mu}A^{\mu} - \lambda \left(  \partial _{\mu}A^{\mu}\right)
+ \frac{1}{2} \partial_{\mu}\phi\partial^
{\mu}\phi - \frac{m^{2}}{4\lambda} \phi ^{2}\ .\eqno(6)$$
This is the standard St\"uckelberg form of the Lagrangian for a
massive Abelian gauge theory! The BRST charge of this model for
the symmetry (4) is given by

$$Q_{BRST} = \int d^{3}x B\partial _{0}c - \partial _{0}Bc =
i \sum_{k}\left( c^{\dag}_{k}B_{k} - B^{\dag}_{k}c_{k} \right) \ ,
\eqno(7)$$
where B is the auxiliary field that linearize the gauge-fixing term
and c denotes the ghost field [11]. The property that Abelian ghosts
decouple imply that the state vector space $V$ can be decomposed
into a direct product $V=V'\otimes V_{FP}$, where the $V_{FP}$
contains only ghost fields and all other fields belong to $V'$.
The physical state condition, $Q_{BRST}|phys>= 0$, takes in
our case the form

$$B_{k}|phys> = 0 \ , for\ all\ k . \eqno(8)$$
When one combine this with the B-field equation of motion, one gets:

$$\left( \partial _{\mu} A^{\mu} - \frac{m}{\lambda}\phi \right)_{k}
|phys>=0 \eqno(9)$$
which is precisely the St\"uckelbeg physical state condition. Let us
notice that, although the gauge fixing term breaks the gauge
symmetry (4), the Lagrange density (6)
is still invariant with respect to (4) if

$$\Box \alpha + \frac{m^{2}}{\lambda} \alpha = 0 \ . \eqno(10)$$
This explains the source of the "extra" symmetry of the
St\"uckelberg model: the gauge fixing condition allows it. Of
course, other gauge fixing conditions are also possible. they will
give us other possible forms of a massive Abelian gauge field model.
It is obvious that the condition $\phi = 0$ (unitary gauge) leads
to (1).

\section{Non-Abelian case}

\ \ \ The non-Abelian massive gauge field has the following Lagrange
density:

$${\it L}=-\frac{1}{2} TrF_{\mu \nu}F^{\mu \nu} + m^{2}Tr\left(
A_{\mu}A^{\mu}\right) \ , \eqno(11)$$
where

$$F^{a}_{\mu \nu} =\partial _{\mu}A_{\nu}^{a} - \partial _
{\nu}A_{\mu}^{a} + gf^{abc}A_{\mu}^{b}A_{\nu}^{c} \ .$$
To generalize our construction to the non-Abelian case,
let us perform the field-enlarging transformation

$$A_{\mu}= U^{\dag}A'_{\mu} U - \frac{i}{g}U^{\dag}\partial _{\mu} U
\ , \eqno(12)$$
where the field U takes values in the adjoint (unitary)
representation of the gauge group. This results in (as before we
drop the prime sign over the gauge field)

$${\it L}=-\frac{1}{2}TrF_{\mu \nu}F^{\mu \nu} + m^{2}Tr\left(
A_{\mu}A^{\mu}\right) - 2\frac{im^{2}}{g}Tr\left( \partial _{\mu}
UU^{\dag}A^{\mu}\right) - \frac{m^{2}}{g^{2}}Tr\left(
U^{\dag}\partial _{\mu}UU^{\dag}\partial ^{\mu}U\right) \ .
 \eqno(13)$$
It is convenient to write the $U$ field as

$$U\left( x\right) = exp\left( \frac{ig}{m}\phi^{a}\left( x\right)
T^{a}\right) \ , $$
where $T^{a}$ denotes the Lie algebra generators of the gauge group.
The Eq. (13) can be then rewritten to

$${\it L}=-\frac{1}{2}TrF_{\mu \nu}F^{\mu \nu} + m^{2}Tr\left(
A_{\mu}A^{\mu}\right) + 2m Tr\left( \partial _{\mu} \phi
A^{\mu}\right) + Tr\left( \partial _{\mu} \phi \partial
^{\mu}\phi\right)\ . \eqno(14)$$
As in the Abelian case, this Lagrange density is invariant with
respect to the following gauge transformations

$$\delta\phi ^{a}\left( x\right) =\alpha^{a} \left( x\right)
\eqno(15a)$$

$$\delta A^{a}_{\mu}\left( x\right) = -\left( D_{\mu}\alpha
\right)^{a}\left( x\right) \ , \eqno(15a)$$
where $D_{\mu}$ denotes the covariant derrivative. To quantize the
model we have to choose a gauge condition. The gauge-
fixing condition

$${\it L}_{gf}=-\lambda Tr\left( \partial _{\mu}A^{\mu} - \frac{m}
{\lambda}\phi \right) ^{2} \eqno(16)$$
leads to te Lagrange density (we omit ghost fields)

$${\it L}=-\frac{1}{2}TrF_{\mu \nu}F^{\mu \nu} + m^{2}Tr\left(
A_{\mu}A^{\mu}\right) + \lambda Tr\left( \partial _{\mu} A^{\mu}
\right)^{2} + Tr\left( \partial _{\mu} \phi \partial ^{\mu}\phi\right)
-\frac{m^{2}}{\lambda}Tr\phi ^{2} \eqno(17)$$
which is the standard St\"uckelbeg's one. Other gauge conditions
provide us with more sophisticated forms of the massive non-Abelian
gauge field Lagrangians. The BRST charge, due to the presence of a
more complicated ghost sectors, has not such an obvious
interpretation as in the Abeliam case, but it still contains the
condition that removes the scalar component of $A_{\mu}$. Indeed, we
can write

$$Q_{BRST}= c^{a}\left(\partial _{\mu} A^{\mu} -\frac{m}
{\lambda}\phi \right) _{a}
- \frac{i}{2}f^{c}_{ab}c^{a}c^{b}\pi _{c} \ , \eqno(18)$$
where $c^{a}$ and $\pi _{a}$ are the ghosts and their canonical
conjugate fields. The we have

$$Q_{BRST}|\psi> = c^{a}G_{a}\psi ^{\left( 0\right)}
+ \frac{1}{2}c^{a}c^{b}
\left[ G_{a}\psi ^{\left( 1\right)}_{b} - G_{b}\psi ^{\left( 1\right)}
_{a} -if^{c}_{ab}\psi ^{\left( 1\right)}_{c}\right] + \dots
\ ,\eqno(19)$$
where $G=\left(\partial _{\mu} A^{\mu} -\frac{m}{\lambda}\phi
\right)$ and $\psi ^{\left( i\right)}$ denotes the i-ghost-level
component of
the state

$$\psi = \sum_{k=0}^{k=n} \frac{1}{k!} c^{a_{1}}\dots c^{a_{k}}\psi
^{\left( k\right)}_{a_{1}\dots a_{k}}\ . $$
We see that the physical state condition $Q_{BRST}|phys>=0$ leads to

$$\left(\partial _{\mu} A^{a\mu} -\frac{m}{\lambda}\phi
^{a}\right)\psi ^{\left( 0\right)} =0 \ . \eqno(20)$$
Unfortunately, we cannot ensure that there are no physical states
comming from higher ghost level: at least academic examples of such
theories can be given [12]. However, Yang-Mills
theories seem to be safe and the BRST-
physical-state condition forces the St\"uckelberg one [12, 13].

\section{Applications}

\ \ \ In this Section we would like to describe two possible
applications of the described approach to the St\"uckelberg
formalism. First, we shall generalize the approach to
the case of a vector field
with a non-Yang-Mills types of couplings, that are often introduced
while discussing possible deviation from the
orthodox standard model of the
electroweak unification. Then we shall consider the anomalous $U(1)$
chiral gauge theory and show the St\"uckelberg formalism is
related to the Batalin-Fradkin-Vilkovisky procedure.

\subsection{Vector field with non-Yang-Mills types of couplings}

\ \ \ Very often, one has to use an effective Lagrangian as a low
energy approximation to a yet not known theory.. For example, such
considerations are important for analysing the possible existence of
anomalous weak vector bosons couplings [14-19]. One can impose olny
the conditions of Lorentz and $U(1)_{em}$ invariance [14,15] on such
an effective Lagrangian. It is also possible to require invariance
with respect to the $SU_{L}(2)\otimes U_{Y}(1)$ but with the
$SU_{L}(2)$ gauge symmetry nonlinearly realized [16-19].
We would like to show that
by using the St\"uckelbeg formalism these models are related.
Let us suppose that the Lagrangian density

$${\it L}\left( F_{\mu \nu},A_{\mu}, W^{\pm}_{\mu}, Z_{\mu},\psi _{i}
\right) \eqno(21)$$
is a general Lagrangian whose form is constrained only by the
requirement of invariance with respect to the
Lorentz and the $U_{em}(1)$-gauge symmetries.
$W^{\pm}_{\mu}$ and $Z_{\mu}$ denote field mediating
weak inteactions. $A_{\mu}$
is the photon. $F_{\mu \nu}$ denotes the vector field
kinetic term and $\psi _{i}$
all matter fields. The field-enlarging transformation

$$\frac{1}{2}\pmatrix{g_{Z}Z_{\mu}& \sqrt{2}g_{W}W_{\mu}^{+}\cr
\sqrt{2}g_{W}W_{\mu}^{-} & g_{Z}Z_{\mu}\cr}=W_{\mu}
\ \ \rightarrow\ \ W_{\mu}'=U^{\dag}W_{\mu}U - iU^{\dag}\partial
_{\mu}U  \eqno(22)$$

$$\psi \ \ \ \rightarrow\ \ \ R\left( U\right) \psi \ , \eqno(23)$$
where $R$ denotes the appropriate matter field representation
and $U^{\dag}U=1$, introduces a non-linearly realized $SU_{L}(2)$
gauge symmetry to the model. The condition $U^{\dag}U=1$ removes the
physical scalar particle from the spectrum. Effectively,
the transformation (22) can be realized by the substitution

$$g_{W}W_{\mu}^{\pm}\ \ \rightarrow \ \  tr\left[ \tau
^{\pm}W'\right] \eqno(24a)$$

$$g_{Z}Z\ \ \ \rightarrow \ \ tr\left[ \tau ^{\pm}W'\right] \ ,
\eqno(24b)$$
where $\tau ^{\pm}=\frac{1}{2}(\tau _{1} \pm i\tau _{2}$ and
$\tau _{i}$ tenote the $SU(2)$ generators. Note, that if we do not
perform (23) then the matter fields are gauge invariant. So, in fact,
we have two types of gauge symmetry at our disposal. As before,
various gauge fixing conditions leads to different
representations of the model.
This generalizes the considerations presented in [14]. Note, that the
possible cut-offs dependence of the results makes the above
considerations quite non-trivial.

\subsection{Anomalous chiral U(1) gauge theory}

\ \ \ Let us consider the following Lagrange density

$${\it L}= -\frac{1}{4} F_{\mu \nu}F^{\mu \nu} + \frac{m^2}{2}
A_{\mu}A^{\mu}+ i{\bar \psi}_{L}\gamma ^{\mu}\left( \partial
_{\mu} + i gA_{\mu}\right) \psi _{L} \ ,\eqno(25)$$
where$\psi _{L}= \frac{1}{2} (1-\gamma ^{5})\psi $ is the
left-handed Weyl field. The transformation (2) leads to the
following St\"uckelberg Lagrangian

$${\it L}= -\frac{1}{4} F_{\mu \nu}F^{\mu \nu} + \frac{m^2}{2}
\left( A_{\mu} + \partial _{\mu} \theta \right)^{2} +
i{\bar \psi}_{L}\gamma ^{\mu}\left( \partial
_{\mu} + i gA_{\mu} + ig\partial _{\mu}\theta \right) \psi _{L}
\ .\eqno(26)$$
The Fujikawa method [21] can be used to derive the equality
(the path integral is understood)

$$g{\bar \psi _{L}}\gamma ^{\mu} \partial _{\mu} \theta \psi _{L}=
\frac{g^{3}}{32 \pi ^{2}} \epsilon ^{\mu \nu \rho \sigma}
\theta F_{\mu \nu} F_{\rho \sigma}\ .\eqno(27)$$
So finaly, we have

$${\it L}= -\frac{1}{4} F_{\mu \nu}F^{\mu \nu} + \frac{m^2}{2}
\left( A_{\mu} + \partial _{\mu} \theta \right)^{2} +
i{\bar \psi}_{L}\gamma ^{\mu}\left( \partial
_{\mu} + i gA_{\mu}\right) \psi _{L}
- \frac{g^{3}}{32 \pi ^{2}} \epsilon ^{\mu \nu \rho \sigma}
\theta F_{\mu \nu} F_{\rho \sigma}\ .\eqno(28)$$
Here, the last term is the result of the anomalous transformation
of the fermionic determinat. It depends on the spacetime dimension
[9, 10, 20, 21]. Now, it is obvious that the St\"uckelberg formalism
have to be put in force as a field-enlarging transformation. The
addition of the scalar deegrees of freedom alone
would neglect the last term in (28) and the symmetry would
not be restored. The same Lagrangian can be obtained by the
BFV quantization procedure [4,5] (plus the gauge-fixing and ghosts
sectors). To get the orthodox form of the St\"uckelberg
Lagrangian in the BFV formalism one has to choose the correct
gauge condition [6]. In our approach, when the
additional symmetry is explicitely introduced, there is
full analogy between the St\"uckelberg
scalar field and the BFV field. Different gauge conditions result
(equivalent) representations: no special gauge is required.
The explicit form of the additional symmetry allows imediately to
answer the question [6]
why the simultaneous apprerance of both the kinetic term
of the scalar
field $\theta$ and the Wess-Zumino term requires the
presence of the
gauge field mas term. The answer is: the mass term is necessary
because it compesates the transformaton of the scalar field kinetic
term. Otherwise the symmetry would be broken.

\section{Concluding remarks}

\ \ \ We have shown that the St\"uckelberg formalism can be
regarded as a field-enlarging transformation that introduces
an additional gauge symmetry to the model. Such transformation
does not influence the S-matrix because it is a point
transformation. The well known theorems concernig point
transformations imply this [22]. If one fully
explores the BRST structure of the model
one gets that the St\"uckelberg physical state condition is exactly
the requirement that the BRST charge anihilates physical states.
It is also possible  to visualize direct analogies with the
Batalin-Fradkin-Vilkovisky quantization procedure. The St\"ukelberg
approach allows to keep
track of additional symmetries. This is not always possible in
the abstract formulation. The origin of the anitfields can be undrestood
in an analogous way [24]. The
formalism can be also used to analyse the bosonisation phenomenon [9, 23]
and quantization of anomalous chiral theories [10]. Wide application
of the formalism in the effective Lagrangian models, along the lines
can be anticipated [14].

\ \ \ {\bf Acknowledgements}. The author would like to
thank Prof. R. K\"
ogerler and dr K. Ko\l odziej for stimulating and helpful
discussions. This work has been supported
in part by the Alexander
von Humboldt Foundation and the Polish Committee for
Scientific Research
under the contract KBN-PB 2253/2/91.

\newpage
\subsection*{\ \ References}

\newcounter{bban}

\begin{list}
{[\arabic{bban}]}{\usecounter{bban}\setlength{\rightmargin}
{\leftmargin}}

\item J. Alfaro and P. H. Damgaard, Ann. Phys. (NY) 202 (1990) 398;
CERN-preprint CERN-TH6455/92 (1992).
\item A. Hosoya and K. Kikkawa, Nucl. Phys. B101 (1975) 271.
\item E. St\"uckelberg, Helv. Phys. Acta 11 (1938) 299.
\item E.S. Fradkin and G.A. Vilkovisky, Phys Lett. B55 (1975) 224.
\item I.A. Batalin and G.A. Vilkovisky, Phys. Lett. B69 (1977) 309.
\item T. Fujiwara, Y. Igarashi and J. Kubo, Nucl. Phys.
B 341 (1991) 695.
\item N.R.F. Braga and H. Montani, Phys. Lett. B264 125.
\item Yong-Wang Kim et al., Phys. Rev. D46 (1992) 4574.
\item J. S\l adkowski, Phys. Lett. B296 (1992) 361 .
\item J. S\l adkowski, Bielefeld Univ. preprint BI-TP 93/07 (1993).
\item T. Kugo and I. Ojima, Progr. Theor. Phys. Suppl. 66 (1979) 1.
\item W. Kalau, NIKHEF-PhD thesis (1992).
\item J.W. van Holten, Nucl. Phys. B339 (1990) 158.
\item C.P. Burgess and D. London, Montreal Univ. preprint
UdeM-LPN-TH-83 (1992).
\item W. Bernreuther, U. L\"ow, J.P. Ma and O. Nachtmann,
Zeit. Phys. C43 (1989) 117.
\item G. Valencia and A. Soni, Phys. Lett. B263 (1991) 517.
\item R. Peccei, S peris and X. Zhang, Nucl. Phys. B349 (1991) 305
\item G. Cvetic and R. K\"ogerler, Nucl. Phys. B 328 (1989) 342 and
b353 (1991).
\item C. Callanm, S. Coleman, J. Wess and B. Zumino, Phys. Rev. 177
(1969) 2239 and 2247.
\item R. Jackiw, in Relativity, Groups and Topology II, edited by
B. DeWitt and R. Stora (North Holland, Amsterdam 1984), p.224.
\item K. Fujikawa, Phys. Rev. D21 (1980) 2848.
\item R. Flume, Comm. Math. Phys. 40 (1975) 49; CERN preprint
TH2144 (1976).
\item P. H. Damgaard, H. B. Nielsen and R. Sollacher, Nucl. Phys.
B385 (1992) 227.
\item J. Alfaro and P. H. Damgaard, CERN-preprint CERN-TH6788/93
(1993).

\end{list}

\end{document}